\documentclass[superscriptaddress,twocolumn,showpacs,amsmath,amssymb]{revtex4}
\usepackage{graphicx}

\renewcommand{\vec}[1]{\mbox{\boldmath $#1$}}

\begin{document}

\title{Three-body model calculations for 
$N=Z$ odd-odd nuclei with $T=0$ and $T=1$ pairing correlations}
\author{Y. Tanimura}
\affiliation{ 
Department of Physics, Tohoku University, Sendai 980-8578,  Japan} 

\author{H. Sagawa}
\affiliation{
Center for Mathematics and Physics,  University of Aizu, 
Aizu-Wakamatsu, Fukushima 965-8560,  Japan}
\affiliation{
RIKEN Nishina Center, Wako 351-0198, Japan}
%E-mail: sagawa@u-aizu.ac.jp

\author{K. Hagino}
\affiliation{ 
Department of Physics, Tohoku University, Sendai 980-8578,  Japan} 
%E-mail:hagino@nucl.phygs.tohoku.ac.jp

\begin{abstract}
We study the interplay between the isoscalar ($T=0$) and isovector ($T=1$) 
pairing correlations 
in $N=Z$ odd-odd nuclei from $^{14}$N to $^{58}$Cu by using three-body model 
calculations.  
The strong spin-triplet $T=0$ pairing correlation dominates 
in the ground state of $^{14}$N, $^{18}$F, $^{30}$P, and $^{58}$Cu 
with the spin-parity $J^{\pi}=1^+$, which can be well reproduced by the present 
calculations. 
The magnetic dipole and Gamow-Teller transitions are found to be strong 
in $^{18}$F and $^{42}$Sc 
as a manifestation of SU(4) symmetry in the spin-isospin space.  
We also discuss the spin-quadrupole transitions 
in these nuclei. 
\end{abstract}

\pacs{21.10.-k,23.20.-g,21.60.Cs}

\maketitle
\section{Introduction}
The pairing correlation is one of the most remarkable effects in 
nuclear physics. It appears 
in many properties of nuclei, including 
odd-even mass staggering, as well as 
the large energy gap between the first excited 
and the ground states in even-even nuclei compered to 
odd-even nuclei.  
In literature, 
the spin-singlet $T=1$ pairing has been mainly discussed in nuclear physics, 
since the large spin-orbit splitting prevents to couple 
a spin-triplet ($T=0$, $S=1$) pair in the ground state \cite{Bertsch2012,Sagawa2013}.  
Another reason for this is 
that the large neutron-excess along the stability line of 
the nuclear chart suppresses the proton-neutron pairing. 
A recent availability of radioactive beams 
has opened up an opportunity 
to measure structure properties 
of unstable nuclei along the $N=Z$ line, strongly enhancing 
a possibility to measure new properties of nuclei such as pairing 
correlations related with the spin-triplet $T=0$ pairing. 
It is thus quite interesting and important to study the competition between the spin-singlet $T=1$ and the spin-triplet $T=0$ pairing 
interactions in odd-odd $N=Z$ nuclei and 
seek an experimental evidence for 
the competition in the spins of low-lying states. 
In this paper, we focus our study in $sd$- and $pf$- shell nuclei, 
in which  the ground state spins and spin-isospin transitions are observed. 
In order to study the ground state and the low-lying 
excited states in odd-odd $N=Z$ nuclei in these mass regions, 
we apply a three-body model with a density-dependent 
contact interaction between the valence neutron and proton.  

The paper is organized as follows. 
In Sec. II, we explain the three-body model employed in the present study. 
In Sec. III, we present the results of the calculations and discuss the 
ground state properties of 
odd-odd $N=Z$ nuclei. 
We also discuss the magnetic moments, the magnetic dipole transitions, 
the isovector spin-quadrupole transitions, and the Gamow-Teller transitions in these nuclei. 
We summarize the paper in Sec. IV. 

\section{Model}

We first describe the model Hamiltonian for $N=Z$ nuclei, assuming 
the core+ $p + n$ structure \cite{Tanimura12}.   This model is based on the three-body model for describing the properties of 
 Borromean nuclei such as $^{11}$Li and $^{6}$He 
\cite{BeEs91,EsBeH97, HS2005}.
In the rest frame of the three-body system, 
the model Hamiltonian  is given by 
\begin{eqnarray}
H&=&\frac{\vec{p}_p^2}{2m}+\frac{\vec{p}_n^2}{2m}+V_{pC}(\vec{r}_p)+V_{nC}(\vec{r}_n) \nonumber \\
&&+V_{pn}(\vec{r}_p,\vec{r}_n)+\frac{(\vec{p}_p+\vec{p}_n)^2}{2A_{C}m},
\label{eq:H}
\end{eqnarray}
where $m$ is the nucleon mass and $A_C$ is the mass number of the core nucleus. 
$V_{pC}$ and $V_{nC}$ are the mean field potentials 
for the valence proton and 
neutron, respectively, 
generated by the core nucleus. 
These are given as 
\begin{eqnarray}
V_{nC}(\vec{r}_n)(r)=V^{(N)}(r_n),\ V_{pC}(\vec{r}_p)=V^{(N)}(r_p)+V^{(C)}(r_p), 
\label{eq:pot}
\end{eqnarray}
where $V^{(N)}$ and $V^{(C)}$ are the nuclear and the Coulomb parts, respectively. In Eq. (\ref{eq:H}), $V_{pn}$ is the pairing 
interaction between the two valence 
nucleons. 
For simplicity, we neglect in this paper 
the recoil kinetic energy of the core nucleus, that is, 
the last term in Eq. (\ref{eq:H}). 

The nuclear part of the 
core-valence particle interaction, Eq. (\ref{eq:pot}), is taken to be
\begin{equation}
V^{(N)}=v_0f(r)+v_{ls}\frac{1}{r}\frac{df(r)}{dr}(\vec{l}\cdot\vec{s}), 
\label{vnC}
\end{equation}
where $f(r)$ is a Fermi function defined by $ f(r)=1/(1+\exp[(r-R)/a])$. 
For $^{18}$F nucleus, as in Ref. \cite{Tanimura12}, we set $v_0=-49.21$ MeV and $v_{ls}=21.6$ MeV$\cdot$fm$^2$. 
For the 
other nuclei, we adjust $v_0$ so as to reproduce the neutron separation energies, while $v_{ls}$ is kept 
constant for all the nuclei considered in this paper. 
The radius and the diffuseness parameters are set to be 
$R=1.27A_C^{1/3}$ fm and $a=0.67$ fm, respectively.
The Coulomb potential $V^{(C)}$ in the proton mean field potential 
is generated by a uniformly 
charged sphere of radius $R$ and charge $Z_Ce$, 
where $Z_C$ is the atomic number of the core nucleus. 
We use a contact interaction 
between the valence 
neutron and proton, $V_{np}$, given as\cite{Tanimura12}, 
\begin{eqnarray}
V_{np}(\vec{r}_1,\vec{r}_2)&=&\hat{P_s}v_s\delta(\vec{r}_1-\vec{r}_2)
\left[1+x_s\left(\frac{\rho(r)}{\rho_0}\right)^{\alpha}\right]  \nonumber  \\
&&+\hat{P_t}v_t\delta(\vec{r}_1-\vec{r}_2)
\left[1+x_t\left(\frac{\rho(r)}{\rho_0}\right)^{\alpha}\right],
\label{eq:pairing}
\end{eqnarray}
where $\hat{P}_s$ and $\hat{P}_t$ are the projectors onto the 
spin-singlet and spin-triplet channels, respectively: 
\begin{eqnarray}
\hat{P}_s=\frac{1}{4}-\frac{1}{4}{\boldsymbol \sigma}_p\cdot{\boldsymbol \sigma}_n,\ 
\hat{P}_t=\frac{3}{4}+\frac{1}{4}{\boldsymbol \sigma}_p\cdot{\boldsymbol \sigma}_n. 
\end{eqnarray}
In each channel 
in Eq. (\ref{eq:pairing}), 
the first term 
corresponds to the interaction in vacuum while 
the 
second term takes into account the medium effect through the density 
dependence. Here, the core density is assumed to be 
a Fermi distribution of the same radius and diffuseness as 
in the core-valence particle interaction, Eq. (\ref{vnC}).  
The strength parameters, $v_s$ and $v_t$, are determined 
from the proton-neutron scattering length as \cite{EsBeH97}
\begin{eqnarray}
v_s&=&\frac{2\pi^2\hbar^2}{m}\frac{2a^{(s)}_{pn}}{\pi-2a^{(s)}_{pn}k_{\rm cut}},\\
v_t&=&\frac{2\pi^2\hbar^2}{m}\frac{2a^{(t)}_{pn}}{\pi-2a^{(t)}_{pn}k_{\rm cut}}, 
\label{eq:v0pair}
\end{eqnarray}
where $a^{(s)}_{pn}=-23.749$ fm and $a^{(t)}_{pn}=5.424$ fm \cite{KoNi75} are 
the empirical p-n scattering lengths in the spin-singlet and spin-triplet 
channels, respectively. 
$k_{\rm cut}$ is the momentum cut-off 
introduced in treating the 
delta function, 
which is related with the cutoff energy 
as 
$E_{\rm cut}=\hbar^2k_{cut}^2/m.$ 
The strengths $v_s$ and $v_t$ determined from 
the scattering lengths depend on the cutoff energy, %% which is taken to be 
$E_{\rm cut}$, %%=20 MeV. 
as will be discussed in Sec. III. 
The three parameters $x_s,\ x_t$, and $\alpha$ in the density-dependent terms in 
Eq. (\ref{eq:pairing}) are determined so as to reproduce energies of the ground ($J^\pi=1^+$), 
the first excited ($J^\pi=3^+$), and the second excited ($J^\pi=0^+$) 
states in $^{18}$F with 
respect to the three-body threashold (See also Ref. \cite{Tanimura12}). 
The density  $\rho(r)/\rho_0$ is replaced by a Fermi function $f(r)$ hereafter.

\begin{figure}
\begin{center}
\includegraphics[scale=.33 ,angle=-90]{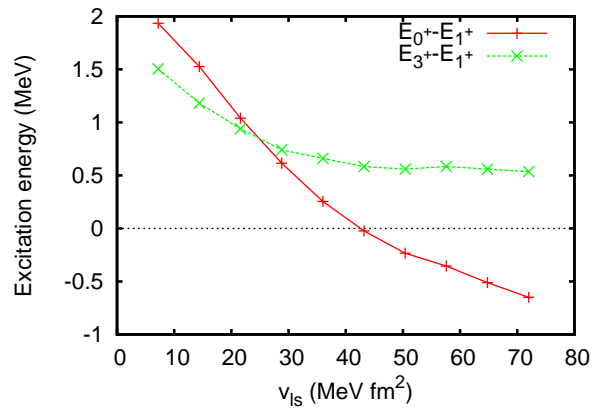}
\end{center}
\caption{(Color online) The excitation energies of the first 0$^+_1$ 
and the first 3$^+_1$ states in $^{18}$F obtained with the three-body 
model as a function of the spin-orbit strength $v_{ls}$ in the mean-field 
potential. The excitation energies are 
measured from the energy of the first 1$^+_1$ state.}
\label{fig:Ex-18F}
\end{figure}

\begin{figure}
\begin{center}
\includegraphics[scale=.33 ,angle=90]{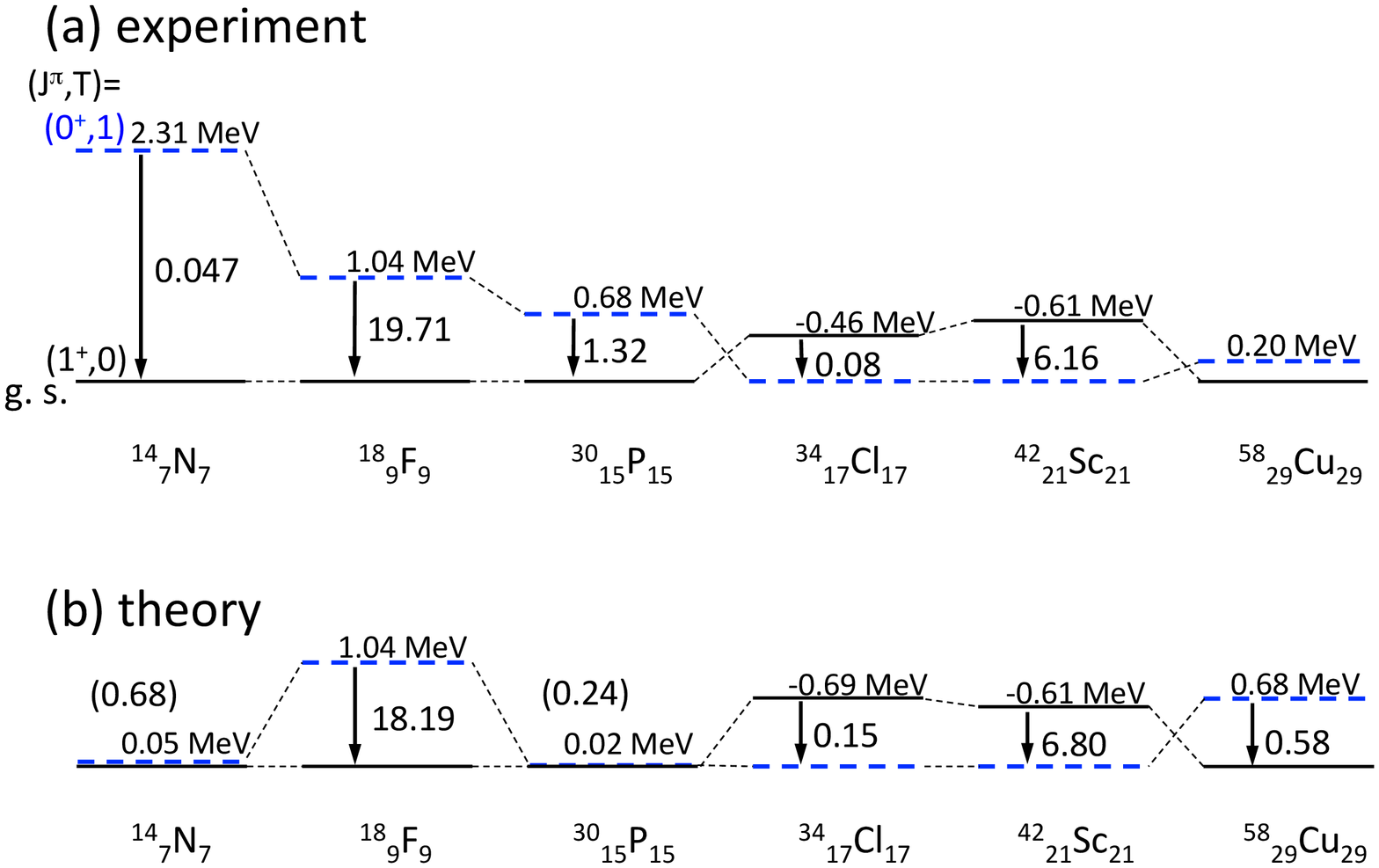}
\end{center}
\caption{(Color online) The energies of the first 0$^+_1$ and the 
first 1$^+_1$ states in $N=Z$ nuclei. The upper panel (a) shows experimental data and the lower panel (b) corresponds to calculated results.
The values with the arrows 
show the  transition probabilities for the  
magnetic dipole transitions, $B(M1)$ (the calcurated values are shown in the brakets for $^{14}$N and $^{30}$P). 
The experimental data are taken from Ref. \cite{exp}.}
\label{fig:levelN=Z}
\end{figure}

The Hamiltonian (\ref{eq:H}) is diagonalized in the valence two-particle 
model space.  The basis states for this are 
given by a product of proton and neutron single 
particle states with the single particle energy $\epsilon^{(\tau)}$,  
which are obtained with the single-particle potential $V_{\tau C}$ in 
Eq. (\ref{eq:H}) ($\tau=p$ or $n$). 
To this end, 
the single-particle continuum states are discreized in a large box.  
We include only those states satisfying 
$\epsilon^{(p)}_{\alpha}+\epsilon^{(n)}_{\beta}\leq E_{\rm cut}$.
We use the proton-neutron formalism without antisymmetrization 
in order to take into account the breaking of the isospin symmetry 
due to the Coulomb interaction. 

\section{Results}

The spin-orbit potential in the mean-field potentials 
plays a crucial role 
in determining 
the properties of $T=0$ pairing as discussed 
in Refs. \cite{Bertsch2012,Poves98,Bertsch11}.  
In Fig. \ref{fig:Ex-18F},
we plot the energy differences between 
the first 0$^+$ and 1$^+$ states and between the first 3$^+$ and 1$^+$ states 
%$E_{0^+}- E_{1^+}$ and $E_{3^+}- E_{1^+}$ 
in $^{18}$F 
as a function of the spin-orbit coupling strength $v_{ls}$. 
We use the cutoff energy of $E_{\rm cut}=20$ MeV. 
It is clearly seen in Fig. \ref{fig:Ex-18F} that
the $T=0$ pairing correlations decreases as 
the spin-orbit interaction increases. 
That is, the energy difference $E_{0^+}- E_{1^+}$ decreases 
and eventually the spectrum is reversed 
so that the 0$^+$ state becomes the ground state, 
where the $T=1$ pairing overcomes the $T=0$ pairing.

The calculated spectra for 
$^{14}$N, $^{18}$F, $^{30}$P, $^{34}$Cl, $^{42}$Sc, and $^{58}$Cu nuclei are 
shown in Fig.  \ref{fig:levelN=Z} together with the experimental data.  
The spin-parity for 
the ground state of the nuclei in Fig.  \ref{fig:levelN=Z} are 
$J^{\pi}=1^+$ except for 
$^{34}$Cl and $^{42}$Sc. 
This feature is entirely due to the interplay between the isoscalar 
spin-triplet and the isovector spin-singlet 
pairing interactions in these $N=Z$ nuclei.  
In the present calculations, the ratio between the isoscalar and the 
isovector pairing interactions is $v_t/v_s=1.9$ for the energy cutoff 
of the model space, $E_{\rm cut}=20$ MeV. 
This ratio is somewhat larger than the value $\sim$1.6 obtained in 
Ref. \cite{Bertsch11} from the shell model matrix elements in $p$- and $sd$-shell nuclei. 
For a larger model space with $E_{\rm cut}=30$ MeV, 
the ratio becomes 1.6, but the agreement 
between the experimental data and the calculations somewhat 
worsens quantitatively even though the general feature remains the same.   
It is remarkable that the energy differences 
$\Delta E=E(0^+_1)-E(1^+_1$) are well reproduced in $^{34}$Cl and $^{42}$Sc 
both qualitatively  (the inversion of the 1$^+$  and 0$^+$ states in the 
ground state) and quantitatively (the absolute value of the energy 
difference).  The model description is somewhat poor in $^{14}$N and $^{30}$P
because the cores of these two nuclei are deformed, although 
 the ordering of the two lowest levels are correctly reproduced. 

The probability of the total spin $S=0$ 
and $S=1$ components 
for the $0^+$ and the 1$^+$ states, respectively, 
are listed in Table \ref{tab:1}.  
The total spin $S=0$ and $S=1$ components in two particle configurations 
can be calculated 
with a formula
\begin{eqnarray}
 |(j_{\pi}j_{\nu}) J\rangle=\sum_{L,S}\left\{  \begin{array}{ccc}
   l_{\pi} & l_{\nu}    &L  \\
   s  &  s &   S  \\
   j_{\pi} &  j_{\nu} & J   \end{array} \right \}     \hat{L} \hat{S} \hat{ j_{\pi} }\hat{ j_{\nu} } 
   |(  l_{\pi}  l_{\nu} )LS;J\rangle    \nonumber \\
   \end{eqnarray}
with the $9j$ symbol and a factor $ \hat{L} \equiv\sqrt{2L+1}$.   
For a $ j_{\pi} = j_{\nu} =j=l+1/2$ configuration, 
the $S=0$ and $S=1$ components are given by the factors 
$(j+1/2)/2j$ and  $(j-1/2)/2j$, respectively, 
for $J$=0. 
For a $ j_{\pi} = j_{\nu} =j=l-1/2$ configuration, 
on the other hand, they are 
$(j+1/2)/(2j+2)$ and $(j+3/2)/(2j+2)$ for $S=0$ and $S=1$, respectively. 
Notice that $s_{1/2}^2$ configuration has only $S=0$ component if $J=0$. 
Otherwise, all the two particle states have a 
large mixture of the $S=0$ and $S=1$ components.
In general, the $S=1$ and $S=0$ components are thus 
largely mixed in the wave 
functions of both the ground and the excited states. 
An exception is $^{30}$P. 
In this nucleus, the dominant configuration in the 0$^+$ state 
is $(2s_{1/2}^{\pi}\otimes 2s_{1/2}^{\nu})$, 
which can couple only to the total spin $S=0$. 
On the other hand, in the 1$^+$  state, the dominant 
configuration is 
$(2s_{1/2}\otimes 1d_{3/2})~{T=0}$ which can couple only to 
the total spin $S=1$ 
with the total angular momentum $L=2$.

\begin{table}
\begin{center}
\caption{The energy difference between the 0$^+_1$ and $1^+_1$ states, 
$\Delta E=E(0^+_1)-E(1^+_1$), in $N=Z$ nuclei.  The probabilities of the $S=0$ 
component $P(S=0)$ in the wave functions for the $0_1^+$ state 
are shown in the fourth line. The fifth line shows the probability of the $S=1$ 
component in the 1$^+$ state. 
The probabilities $P(j^{\pi}\otimes j^{\nu})$ for the 
dominant valence shell proton-neutron configuration are 
also given for the 0$^+_1$ and 1$^+_1$ states in the 7th and 8th lines, 
respectively.  
The experimental data is taken from Ref. \cite{exp}.  }
\begin{tabular}{cc|cccccc}
\hline\hline
 & & $^{14}$N & $^{18}$F & $^{30}$P & $^{34}$Cl & $^{42}$Sc & $^{58}$Cu \\
\hline
 $\Delta E$ & exp. & 2.31 & 1.04 & 0.68 & $-0.46$ & $-0.61$ & 0.20 \\
 (MeV) &                 cal. & 0.05 & 1.04 & 0.02 & $-0.69$ & $-0.61$ & 0.68 \\
\hline
 $P(S=0)$ (\%) & $0^+$           & 34.8 & 82.2& 94.8 & 40.7 & 70.5 & 65.4 \\
 $P(S=1)$ (\%) & $1^+$           & 78.3 & 90.1 & 95.8 & 64.3 & 65.7 & 92.1 \\
  \hline
 \multicolumn{2}{c|}{$j$}               & $1p_{1/2}$ & $1d_{5/2}$ & $2s_{1/2}$ & $1d_{3/2}$ & $1f_{7/2}$ & $2p_{3/2}$ \\\hline
$P(j^{\pi}\otimes j^{\nu})$   & $0^+$          & 97.2 & 85.2 & 89.7 & 98.6 & 94.2 & 81.2 \\
(\%) & $1^+$      & 96.4 & 52.1 & 1.1 & 98.4 & 82.7 & 10.0 \\
\hline\hline  \label{tab:1}
\end{tabular}
\end{center}
\end{table}

We next discuss 
the magnetic moment for the $1^+$ state, and
the magnetic dipole transition 
strength 
$B(M1)\downarrow$ 
and 
the isovector spin-quadrupole transition strength 
$B(IVSQ)\uparrow$ 
between $0^+_1$ and $1^+_1$ states.  The symbol $\downarrow (\uparrow)$ means the transition from the excited (ground) to the ground (excited) states.
The magnetic operator is defined as
\begin{equation}
\mu=\langle 1^+|\sum_{i}(g_s(i)\vec{s}_i+g_l(i)\vec{l}_i)|1^+\rangle,
\label{eq:mm}
\end{equation}
where $g_s(i)$ and $g_l(i)$ are the spin and the orbital $g$ factors, 
respectively.  
The reduced magnetic dipole transition probability is given by
\begin{eqnarray}
 &&B(M1:J_i \rightarrow J_f) =  \nonumber \\
&& \left(\frac{3}{4\pi}\right)\frac{1}{2J_i+1}\left|\langle 
J_f||\sum_{i}(g_s(i)\vec{s}_i+g_l(i)\vec{l}_i)||J_i\rangle\right|^2,
\label{eq:m1}
\end{eqnarray}
where the double bar means the reduced matrix element in the spin space.    
We take the bare $g$ factors $g_s(\pi)=5.58 $, $g_s(\nu)=-3.82$, 
$g_l(\pi)=1$, and $g_l(\nu)=0$ for the magnetic moment and the magnetic 
dipole transitions in the 
unit of the nuclear magneton $\mu_N=e\hbar/2mc$. 
The spin-quadrupole transition is defined by
\begin{eqnarray}
 &&B(IVSQ:J_i \rightarrow J_f) =  \nonumber \\
&& \frac{1}{2J_i+1}\left|\langle J_f||\sum_{i}(\tau_z(i)r_i^2[\vec\sigma(i)Y_2(i)]
^{(\lambda=1)})||J_i\rangle\right|^2. 
\label{eq:IVSQ}
\end{eqnarray}

The calculated magnetic moments and the magnetic dipole transitions 
are listed in Table \ref{tab:2} together with the spin quadrupole transitions.
The calculated magnetic moment in $^{14}$N reproduces well the observed one, while the agreement is 
worse 
in $^{58}$Cu.
This is  due to  the fact  that the core of $^{56}$Ni might be  largely broken and the  $f_{7/2}-$hole configuration is mixed in the ground state of $^{58}$Cu 
~\cite{Honma2004}.
The values for $B(M1)$ are also shown in Fig. \ref{fig:levelN=Z}. 
Very strong $B(M1)$ values are found 
both experimentally and theoretically in two of the $N=Z$ nuclei 
in Table \ref{tab:2}, that is, in $^{18}$F and $^{42}$Sc. 
The $B(M1)$ value from 0$^+$ to 1$^+$ in 
$^{18}$F is the largest one so far observed in the entire 
region of nuclear chart. 
We notice that our three-body calculations provide remarkable agreements 
not only for these strong transitions in $^{18}$F and $^{42}$Sc but 
also quenched transitions in the other $N=Z$ nuclei such as in 
$^{14}$N and $^{34}$Cl.

  \begin{table}
\caption{The magnetic moments $\mu$, the 
magnetic dipole transitions, and the 
isovector spin quadrupole transitions in the $N=Z$ nuclei. 
The experimental data of $B(M1)$ values are taken from Ref. \cite{exp}, 
while the data for the magnetic moment are 
taken from Ref. \cite{mmdata}.
The symbol $\downarrow (\uparrow)$ means the transition from the excited (ground) to the ground (excited) states.}
\begin{center}
\begin{tabular}{cc|cccccc}
\hline\hline
 & & $^{14}$N & $^{18}$F & $^{30}$P & $^{34}$Cl & $^{42}$Sc & $^{58}$Cu \\
\hline
%% $J^\pi_{\rm gs}$ & & $1^+$ & $1^+$ & $1^+$ & $0^+$ & $0^+$ & $1^+$\\
$J^{\pi}_{%%\rm 
gs}$ & & $1^+$ & $1^+$ & $1^+$ & $0^+$ & $0^+$ & $1^+$\\
\hline
 $\mu$ ($\mu_N$) & exp. & 0.404 & -- & -- & -- & -- & 0.52 \\
                 & cal. & 0.379 & 0.834 & 0.318 & 0.426 & 0.686 & 0.283 \\
\hline
 $B({%%\rm 
 M}1)\downarrow
 $   ($\mu_N^2$)   & exp. & 0.047 & 19.71 & 1.32 & 0.08 & 6.16 & -- \\
                                & cal. & 0.682 & 18.19 & 0.24 & 0.15 & 6.81 & 0.580 \\
\hline
 $B({%%\rm 
 SQ})\uparrow
 $ (fm$^4$) & cal. & 33.17 & 0.85 & 43.04 & 74.52 & 19.61 & 71.55 \\
%%&&&&&&  
 \hline\hline  \label{tab:2}
\end{tabular}
\end{center}
\end{table}
  
In the case of $^{18}$F, the $0^+$ and $1^+$ states are 
largely dominated by the $S=0$ and $S=1$ 
spin components, respectively, with the orbital angular momentum $l=2$ (see 
Table I). 
Therefore, the two states can be considered as members of 
SU(4) multiplet 
in the spin-isospin space. 
This is the main reason why the $B(M1)$ value
is so large in this nucleus, 
since the spin-isospin operator $g_s^{IV}\vec{s}\tau_z$ 
connects between two states in the same SU(4) multiplet, 
that is, the transition is allowed, 
and the isovector $g-$ factor is the dominant term in Eq. (\ref{eq:m1})
with $g_s^{IV}=(g_s(\nu)-g_s(\pi))/2=-4.70$. 
The configurations in $^{42}$Sc are also similar to 
those in $^{18}$F in terms of SU(4) multiplets, 
although they are dominated by $l=3$ wave functions.  
For $^{14}$N and $^{34}$Cl, 
the $B(M1)$ transitions do not acquire any enhancement, since 
the $S=0$ component in the $0^+$ state 
is suppressed due to the $j=l-1/2$ coupling 
(both the $0^+$ and $1^+$ states have very large $1p_{1/2}^2$ ($1d_{3/2}^2$) configurations in $^{14}$N  ($^{34}$C).) 
These indications for the SU(4) symmetry in $^{18}$F and $^{42}$Sc are 
consistent with the results obtained in Refs. \cite{HaBa89,VoOr93,IsWaBr95}. 
In nuclei $^{30}$P and $^{58}$Cu,  the 1$^+$ state is 
dominated by 1$d_{3/2}2s_{1/2}$ and 
$2p_{3/2}1f_{5/2}$ configurations, respectively, while the 0$^+$ 
state is governed by 
the $2s_{1/2}^2$ and $2p_{3/2}^2$ configurations, respectively. 
Therefore the isovector spin-quadrupole transitions 
are largely enhanced in the two nuclei even though 
the $B(M1)$ is much quenched.  
 
We also calculate 
the Gamow-Teller (GT) strength
\begin{eqnarray}
B(GT:0^+ \rightarrow 1^+) = %\nonumber \\
\frac{g_A^2}{4\pi}\left|\langle1^+||\sum_{i}t_-(i)\vec\sigma(i)||0^+
\rangle\right|^2,
\label{eq:GT}
\end{eqnarray}
where $g_A$ is the axial-vector strength, 
and summarize the results in Table \ref{tab:GT}. 
One can again see the strong GT transition between the 
lowest $0^+$ and $1^+$ states in $A=18$ and 42 systems, 
which exhaust a large portion of the GT sum rule value.  
This can also be interpreted as 
a manifestation of SU(4) symmetry in the wave functions of these nuclei. 
We note here again that the result obtained in Ref. \cite{HaBa89} by an analysis of 
GT transition also implies a good SU(4) symmetry in the $A=18$ system. 
On the other hand, for $^{58}$Cu, 
the GT strength is largely fragmented and 
no strong state in $B(GT)$ is seen near the ground state. 
The experimental data are 
consistent with the calcuted results as can be seen in Table \ref{tab:GT}. 

\begin{table}
\caption{The  Gamow-Teller transition probabilities  from the ground states of $^{18}$O to , $^{42}$Ca and  $^{58}$Ni nuclei, 
in the units of $g_A^2/4\pi$. 
The experimental data are taken from Ref. \cite{GT1} for $^{18}$F,  Ref. \cite{GT2}for $^{42}$Sc and Ref. \cite{GT3} for $^{58}$Cu, respectively .}
\begin{center}
\begin{tabular}{cc|cc}
\hline\hline
 \multicolumn{4}{c}{$^{18}$O $\rightarrow ^{18}$F} \\
\hline 
 \multicolumn{2}{c|}{$E_x$ (MeV)} & \multicolumn{2}{c}{$B(GT)$ ($g_A^2/4\pi$)} \\
\hline
 cal. & (exp.)    &    cal. & (exp.)  \\
\hline
  0.0 & (0.0)      &      2.48 & (3.11$\pm$ 0.03)  \\
  4.79 & (---)    &     0.028 & (---) \\
  6.87 & (---) &       0.036  & (---) \\
\hline
  \multicolumn{4} {c}{$^{42}$Ca $\rightarrow ^{42}$Sc}   \\\hline 
 \multicolumn{2}{c|}{$E_x$ (MeV)} & \multicolumn{2}{c}{$B(GT)$ ($g_A^2/4\pi$)} \\
\hline
 cal. & (exp.)    &    cal. & (exp.)  \\
\hline 
  0.61 & (0.61)   &   1.80 &(2.16  $\pm$ 0.05)  \\
  --- & (1.89)   &    --- &(0.09 $\pm$ 0.01)  \\
  3.71 & (3.69)  &   0.346 & (0.15 $\pm$ 0.02) \\
\hline
 \multicolumn{4} {c}{$^{58}$Ni $\rightarrow ^{58}$Cu}   \\
\hline 
 \multicolumn{2}{c|}{$E_x$ (MeV)} & \multicolumn{2}{c}{$B(GT)$ ($g_A^2/4\pi$)} \\
\hline
 cal. & (exp.)    &    cal. & (exp.)  \\
\hline   
 0.0 & (0.0)   &     0.097 & (0.155  $\pm$ 0.01) \\
 1.24 &(1.05)  &    0.74 & (0.32  $\pm$   0.03)  \\
\hline\hline
\label{tab:GT}
\end{tabular}
\end{center}
\end{table}

\section{summary}
We have studied the properties of the 
lowest 0$^+$ and 1$^+$ states in the odd-odd $N=Z$ nuclei in the 
$sd$- and $pf$- shell region with the three-body model 
with valence proton and neutron and a core.
The ratio between the spin-triplet isoscalar and the spin-triplet 
isovector pairing interactions,  $v_s/v_t$, 
is determined to be 1.9 based on the neutron-proton scattering 
lengths and the energy cut-off of the model space. 
It was pointed out that the energy ordering of the 0$^+$ and 1$^+$ is 
very sensitive to the strength of spin-orbit coupling, i.e., the spin-orbit splitting prevents the strong spin-triplet pairing interactions and 
makes the ground states of $^{34}$Cl and $^{42}$Sc  to have  $J^{\pi}=0^+$. 
The energy differences between the lowest 0$^+$ and 1$^+$ states  are well 
reproduced by our model qualitatively (that is, the inversion of the level 
ordering between the two states)  and quantitatively (that is, the excitation energy). 
It was shown that the calculated wave functions 
of the lowest 0$^+$ and 1$^+$ states in $^{18}$F and $^{42}$Sc have typical features of 
the SU(4) multiplets in the spin-isospin  space and give the strong magnetic dipole transitions strength 
between the $0^+$ and $1^+$ states. The GT transitions from the neighboring even-even $T=1,T_z=1$ 
nuclei $^{18}$O and $^{42}$Ca with the $J^\pi=0^+$ to the 1$^+$ states in the odd-odd $T=0$ nuclei $^{18}$F 
and $^{42}$Sc are also shown to be very strong, exhausting a substantial amount of the GT sum rule. 
The calculated transitions give quantitatively good accounts of 
the observed strong $B(M1)$ and $B(GT)$ values in the two nuclei. 
In the other $N=Z$ nuclei,  $B(M1)$ transitions are rather hindered, while 
the spin-quadrupole transitions are found to be rather strong.

\section*{Acknowledgements}
We would like to thank G.F. Bertsch and B. A. Brown for useful discussions. 
We acknowledge also Y. Fujita for informing us experimental data of Gamow-Teller transitions. 
This work was supported through a grant-in-aid by the JSPS under the program number 24$\cdot$3429 
and the Japanese Ministry of Education, Culture, Sports, Science and Technology by a Grant-in-Aid 
for Scientific Research under program number (C) 22540262. The work of Y.T. was also supported by 
the Japan Society for the Promotion of Science for Young Scientists.


\begin{thebibliography}{99}
\bibitem{Bertsch2012}  G. F. Bertsch, 50 years of nuclear BCS (edited by R. A. Broglia and V. Zelevinsky, World Scientific, 2012).
\bibitem{Sagawa2013}   H. Sagawa , Y. Tanimura and K, Hagino , Phys. Rev. C {\bf 87}, 034310 (2013).
\bibitem{Tanimura12}  Y. Tanimura, K, Hagino and H. Sagawa, Phys. Rev. C {\bf 86}, 044331 (2012). 
\bibitem{BeEs91}G. F. Bertsch and H. Esbensen, Ann. Phys. (NY) {\bf 209}, 327 (1991).
\bibitem{EsBeH97}H. Esbensen, G. F. Bertsch, and K. Hencken, Phys. Rev. C {\bf 56}, 3054 (1997).
\bibitem{HS2005} K. Hagino and H. Sagawa,  Phys. Rev. C {\bf 72}, 044321 (2005).  
\bibitem{KoNi75}L. Koester and W. Nistler, Z. Phys. A {\bf 272}, 189 (1975). 
\bibitem{Poves98}A. Poves and G. Martinez-Pinedo, Phys. Lett. B{\bf 430}, 203(1998).
\bibitem{Bertsch11}  G. F.  Bertsch and Y. Luo,  Phys. Rev. C {\bf 81}, 064320 (2010).  
\bibitem{exp}  Chart of Nuclides,   National Nuclear Data Center (http://www.nndc.bnl.gov/).
\bibitem{mmdata}  N. J. Stone et al., Phys. Rev. C {\bf 77}, 067302 (2008). 
\bibitem{Honma2004} M. Honma, T. Otsuka, B. A. Brown, and T. Mizusaki, Phys. Rev.
{\bf C69}, 034335 (2004)
\bibitem{HaBa89}P. Halse and B. R. Barrett, Ann. Phys. (N. Y.) {\bf 192}, 204 (1989). 
\bibitem{VoOr93}P. Vogel and W. E. Ormand, Phys. Rev. C {\bf 47}, 623 (1993). 
\bibitem{IsWaBr95}P. Van Isacker, D. D. Warner, and D. S. Brenner, 
Phys. Rev. Lett. {\bf 74}, 4607 (1995). 
\bibitem{GT1}   D.R. Tilley, H.R. Weller,   C.M. Cheves and  R.M. Chasteler, Nucl. Phys. A 595, 1 (1995); 
\bibitem{GT2} T. Kurtukian Nieto et al., Phys. Rev. C {\bf 80},  035502 (2009); 
\bibitem{GT3} Y. Fujita et al., EPJ A 13, 411 (2002) and 
Y. Fujita, private communications.


\end{thebibliography}
\end{document}